 \def\a{\alpha}\def\b{\beta}\def\g{\gamma}
  \def\k{\kappa}
  \def\om{\omega}

\def\imo{i}

\def\be{\begin{equation}}
\def\ee{\end{equation}}
\def\bea{\begin{eqnarray}}
\def\eea{\end{eqnarray}}

\def\Re#1{\mathrm{Re}(#1)}
\def\Im#1{\mathrm{Im}(#1)}

\newcommand{\p}[1]{(\ref{#1})}

\documentclass[10pt]{article}
\usepackage{graphicx,rotate}
\usepackage{longtable}
\usepackage{amssymb}
\begin{document}\sloppy
\begin{center}
\textbf{\Large Decay of massive scalar field in a Schwarzschild background}\\

\vspace{5mm}

{\large R.A. Konoplya}\\
Universidade de S\~ao Paulo, Instituto de F\'\i sica\\
Caixa Postal 66318, 05315-970, S\~ao Paulo-SP, Brazil.\\
E-mail: konoplya\_roma@yahoo.com.\\
\vspace{2mm}
and\\
\vspace{2mm}
{\large A.V. Zhidenko}\\
Department of Physics, Dniepropetrovsk National University\\
St. Naukova 13, Dniepropetrovsk  49050, Ukraine.\\
E-mail: Z\_A\_V@ukr.net.
\end{center}
\vspace{2mm}
\begin{abstract}
The decay of massive scalar field in the Schwarzschild black hole
background is investigated here by consideration its quasinormal spectrum.
It has been proved that the so-called $quasi-resonant$ modes,
which are arbitrary long living (purely real) modes,
can exist only if the effective potential is not zero at least at one
of the boundaries of the $R$-region. We have
observed that the quasinormal spectrum exists for all  field masses and proved
 both analytically and numerically that when $n \rightarrow \infty$
the real part of the frequencies approaches the same asymptotical
value ($\ln3/(8\pi M)$) as in the case of the massless field.
\end{abstract}

\vspace{6mm}

At late times, the decay of fields in a black hole background is
dominated by some resonant $quasinormal$ modes (QNMs)
\cite{kokkotas-review}. The interest in QNMs originated from
possibility of detection of gravitational waves from colliding
black holes, especially after creating the new generation of
gravitational antennas. Recently the study of the quasinormal modes
has gained a strong motivation coming from ADS/CFT correspondence,
where the QNMs can be interpreted as the poles of the temperature Green
function in the dual conformal field theory. It is also possible, that
QNMs play an important role in Loop Quantum Gravity. All this
motivated the extensive research of QNMs for different black holes
and different fields (both massive and massless) around them
(see for instance \cite{kucha-mala} and references therein).

Long time ago, when considering the problem of a massive scalar field around
Schwarzschild black hole, A. Starobinskii and I. Novikov
found,  that, when working in the in the frequency domain, the poles
of the Green function in the complex plane are closer to the real axis than in the massless
case. Thus,  it was observed that the ``massive''modes decay more
slowly than the massless ones. Later it was confirmed in a series of
papers, both in frequency domain \cite{1}, \cite{2}, and in  time
domain \cite{4}-\cite{7}, that the greater the mass of the field, the slower
the decay. Then, a natural question is, what will happen at further
increasing of the mass of the field? Can the decreasing decay rate approach zero
leading to existence of  arbitrary long living modes? Which is
high overtone behavior in this case?

In the present paper we are answering these questions by thorough
investigation of the quasinormal behavior corresponding to the decay of a massive scalar field.


{\bf 1. Basic equations}

The massive scalar field in a curved background is governed by the
Klein-Gordon equation:
\begin{equation}
\Box \Phi - m^2  \Phi  = \frac{1}{\sqrt{-g}} \left(g^{ \mu \nu} \sqrt{-g}
\Phi,_{\mu}\right),_{ \nu} - m^2  \Phi = 0.
\end{equation}

After the separation of angular and time variables, the result radial
equation for the Schwarzschild background
\be
ds^2 = f(r)dt^2-\frac{dr^2}{f(r)}-r^2(d\theta^2-\sin^2\theta
d\phi^2), \quad f(r)=1-\frac{2M}{r},
\ee
can be reduced to the following wave-like equation
\be\label{DE}
\left(\frac{d^2}{dr^{*2}}+\omega^2-V(r^*)\right)\Psi(r^*) = 0,
\ee
where
$$dr^* = \frac{dr}{f(r)}, \qquad V(r)=f(r)\left(\frac{l(l+1)}{r^2}+\frac{f'(r)}{r}+m^2\right),$$
and $l=0,1,2,3\ldots$ parameterizes the field angular momentum.

The effective potential has the form of the potential barrier
which  approaches some constant values both at the event horizon and at
spatial infinity. Therefore the QNMs boundary conditions (within the
dominant asymptotic order) have the form:
\be\label{EB}\Psi(r^*)\sim C_{\pm} \exp(\pm \imo \k_{\pm} r^*),\quad
r^*\rightarrow \pm\infty;
\ee
Under these boundary conditions the QNMs of massive  scalar field
were studied in several papers in the WKB approximations
\cite{1},\cite{2}, \cite{3}. Yet, if we need to explore accurate Leaver
method \cite{Leaver}, we have to deal with irregular singular point at infinity.
This implies that one needs to take into consideration the sub-dominant
asymptotic term at infinity:
\be\label{IB}\Psi(r^*)\sim C_+
e^{\imo\chi r^*}r^{(\imo M m^2/\chi)}, \quad
(r,r^*\rightarrow+\infty);\qquad
\chi = \sqrt{\om^2-m^2}.
\ee
(The sign of $\chi$ is to be chosen to stay in the same complex
surface quadrant as $\om$.)

Within the Leaver method we can eliminate the singular factor of the solution \p{DE} that satisfies in-going wave
boundary condition at the horizon  and
\p{IB} at infinity, and expand the remaining part into the Frobenius series that are
convergent in the $R$-region (between the event horizon and the
infinity). The appropriate series are:
\be\label{FS}
\Psi(r) = e^{\imo\chi r}r^{(2\imo M\chi+\imo M
m^2/\chi)}\left(1-\frac{2M}{r}\right)^{-2\imo
M\om}\sum_na_n\left(1-\frac{2M}{r}\right)^n,
\ee

Substituting \p{FS} into \p{DE} we obtain a three-term
recurrent relation for the coefficients $a_n$:
\begin{eqnarray}\label{RR}
&&\a_0a_1+\b_0a_0 = 0;
\\\nonumber &&\a_na_{n+1}+\b_na_n+\g_na_{n-1}=0, \qquad n>0,
\end{eqnarray}
where
\begin{eqnarray}
\nonumber \a_n &=& (n+1)(n+1-4M\imo\om);
\\\label{RC} \b_n &=&\frac{M(\om+\chi)(4M(\om+\chi)^2+\imo(2n+1)(\om+3\chi))}{\chi}-2n(n+1)-1-l(l+1);
\\\nonumber \g_n &=& \left(n-M\imo(\om+\chi)^2/\chi\right)^2.
\end{eqnarray}

Since the series are convergent at infinity, the ratio of the
series coefficients is finite and can be found in two ways:
\be\label{ratio}%
\frac{a_{n+1}}{a_n}= \frac{\g_{n}}{\a_n}\frac{\a_{n-1}}{\b_{n-1}
-\frac{\a_{n-2}\g_{n-1}}{\b_{n-2}-\a_{n-3}\g_{n-2}/\ldots}}-\frac{\b_n}{\a_n}=-\frac{\g_{n+1}}{\b_{n+1}-\frac{\a_{n+1}\g_{n+2}}{\b_{n+2}-\a_{n+2}\g_{n+3}/\ldots}}.
\ee%

Thus we have equation with respect to the eigenvalue $\om$:
\be\label{continued_fraction} \b_n-\frac{\a_{n-1}\g_{n}}{\b_{n-1}
-\frac{\a_{n-2}\g_{n-1}}{\b_{n-2}-\a_{n-3}\g_{n-2}/\ldots}}=
\frac{\a_n\g_{n+1}}{\b_{n+1}-\frac{\a_{n+1}\g_{n+2}}{\b_{n+2}-\a_{n+2}\g_{n+3}/\ldots}},
\ee%
that can be solved numerically.

To improve the convergence of the continued fraction on the right
side of \p{continued_fraction} we used the technique developed by
Nollert \cite{Nollert}. 
He considered the recurrence relation
\be\label{NollertR} R_N =
\frac{\g_N}{\b_N-\frac{\a_N\g_{N+1}}{\b_{N+1}-\a_{N+1}\g_{N+2}/\ldots}}
= \frac{\g_N}{\b_N-\a_NR_{N+1}}.
\ee
Making use of the recurrence relation \p{NollertR} one can find
for large N:
\be\label{NollertSeries} R_N =
C_0+C_1N^{-1/2}+C_2N^{-1}+\ldots
\ee
where
$$C_0=-1,$$
$$C_1=\pm2\sqrt{-\imo M\chi},
\qquad \Re{C_1}>0,$$
$$C_2=\frac{3}{4}+4M\imo\chi+\imo Mm^2/\chi,$$
etc.

The series \p{NollertSeries} converge for $|\chi|/N<A<\infty$,
therefore we can use this approximation for $R_N$ inside the
continued fraction for some $N\gg -\Im\chi\sim n$. In practice, to
find an appropriate $N$ we should increase it until the result of the
continued fraction calculation does not change.

{\bf 2.Quasi-resonances}

Recently  Ohashi and  Sakagami \cite{Ohashi&Sakagami} studied QNMs for
the decay of the massive scalar field and found that there are perturbations
with arbitrary long life when the field mass has special values.
They called these modes $quasi-resonances$.

\begin{figure}
\centerline{\includegraphics{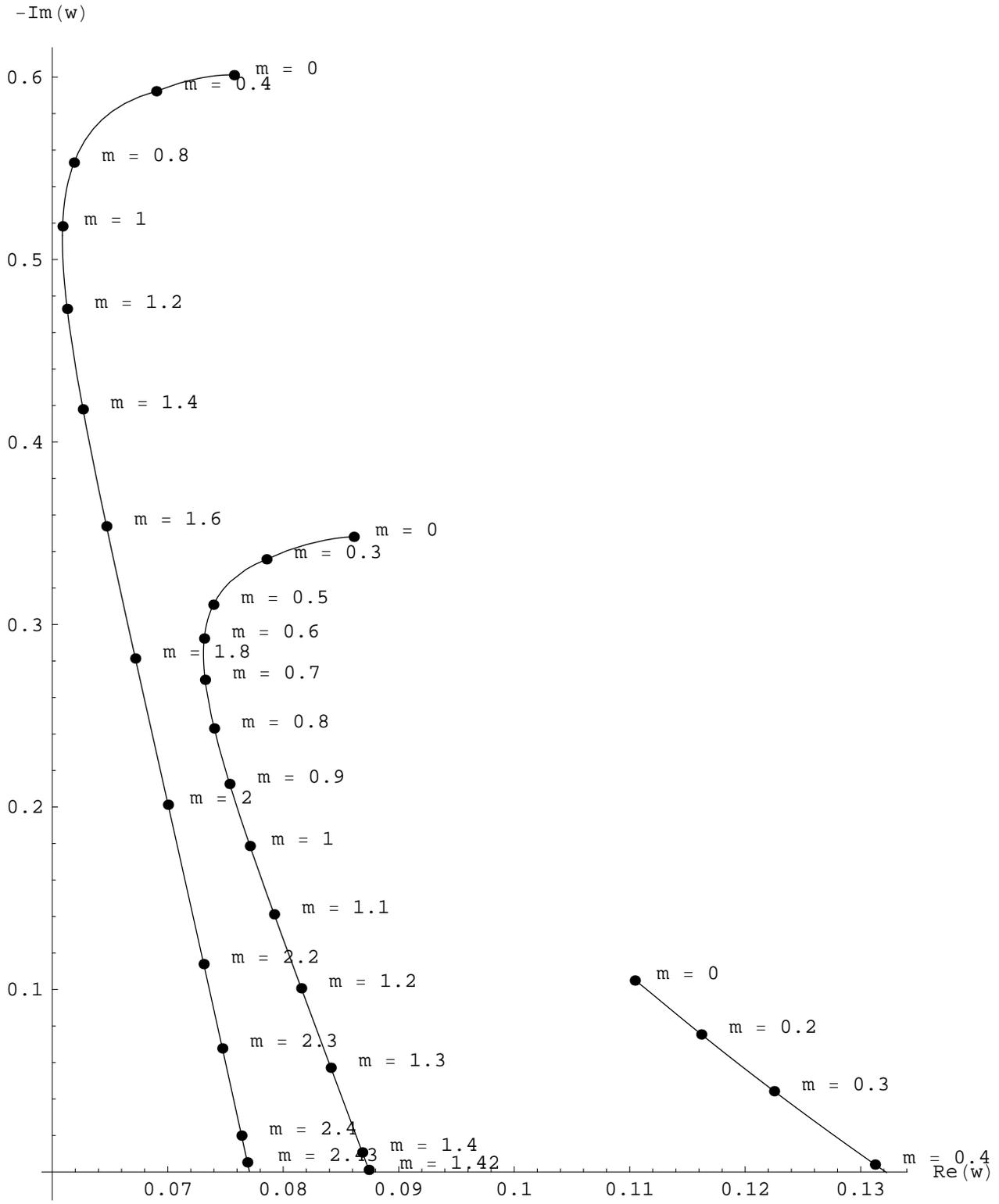}}
\caption{First three overtones of $l=0$ QNMs of the scalar field of
different field masses (M=1)}\label{l=0low}
\end{figure}
\begin{figure}
\centerline{\includegraphics{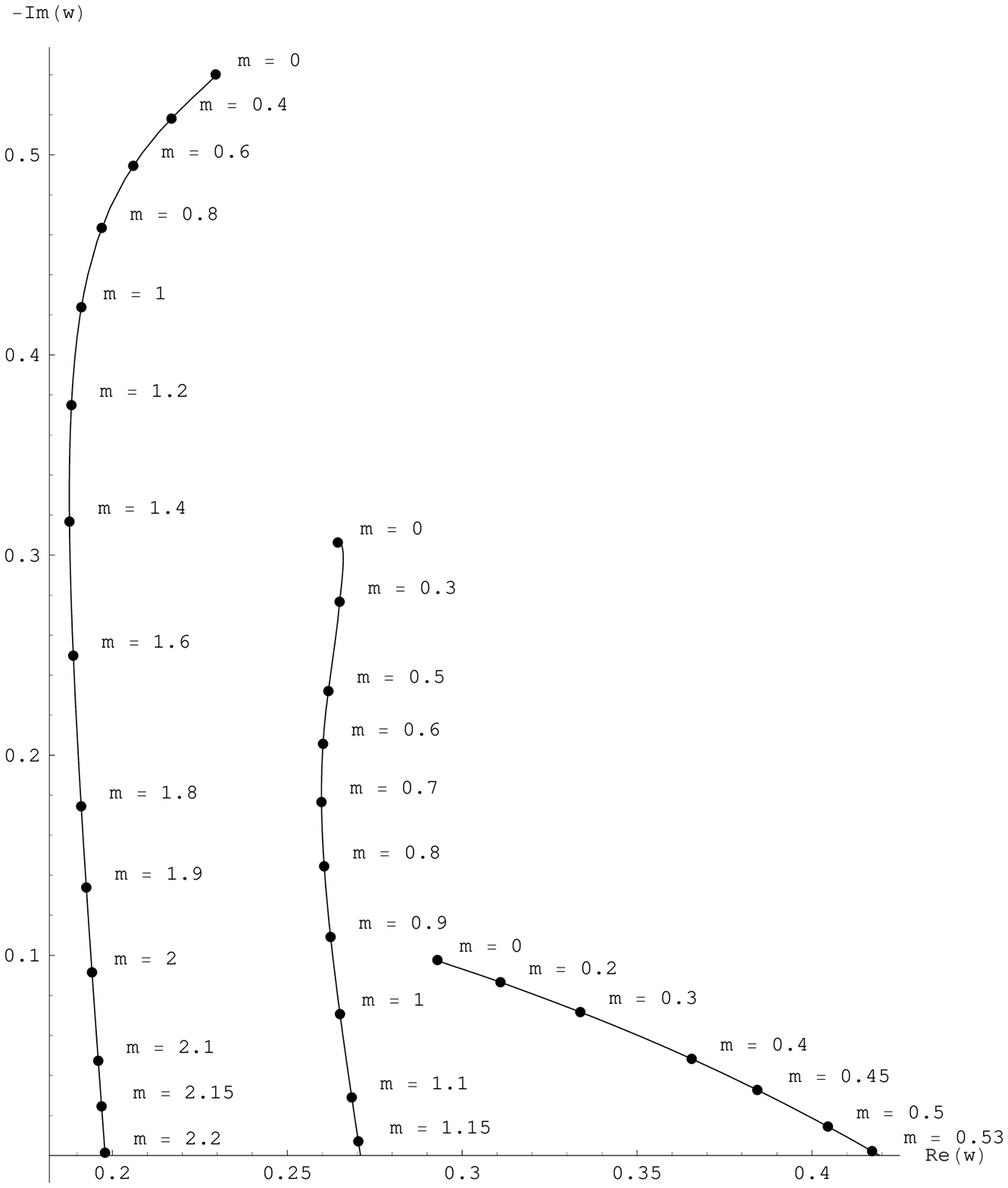}}
\caption{First three overtones of $l=1$ QNMs of the scalar field of
different field masses (M=1)}\label{l=1low}
\end{figure}

The remarkable fact is that in the massive case the purely real
frequencies are not forbidden. Indeed, by multiplying \p{DE} by
$\Psi^*$ and integrating we obtain:
\be\label{EE}
\intop_{-R_1}^{+R_2}\Psi^*\left({\Psi''}+(\omega^2-V)\Psi\right)
dr^*=\omega^2\intop_{-R_1}^{+R_2}|\Psi|^2dr^*-\intop_{-R_1}^{+R_2}(|\Psi'|^2+V|\Psi|^2)dr^*
+\Psi^*\Psi'\bigr|_{r^*=-R_1}^{r^*=+R_2}=0.
\ee
Substituting the boundary conditions for large $R_{1,2}$, we find
that
\bea\Psi^*\Psi'\bigr|_{r^*=-R_1}^{r^*=+R_2} =
\imo\exp(-2\Im{\chi}R_2)R_2^{2M\Im{\chi}m^2/|\chi|^2}\left(\chi|C_+|^2+o\left(\frac{1}{R_2}\right)R_2\right)+\\\nonumber+\imo\exp(-2\Im{\om}R_1)\left(\om|C_-|^2+o\left(\frac{1}{R_1}\right)R_1\right).
\eea
It is easy to see that for the massless case ($\chi=\om$) and
$\Im{\om}=0$, the equation \p{EE} implies
$$\om|C_+|^2=0\qquad \textrm{and} \qquad\om|C_-|^2=0.$$
This means that there is no real frequency in the QNM spectrum
of the massless scalar field.

When the field mass is not zero, the purely real frequencies  $\om$ are
not forbidden because \p{EE} is satisfied when
\be\label{FE}
\Re{\chi}|C_+|^2=0, \qquad \textrm{and} \qquad \om|C_-|^2=0.
\ee
There is no wave falling on the horizon ($C_-=0$) in the latter case,
and there is no energy transmission to infinity ($\Re{\chi}=0$).
Therefore oscillations do not decay. The situation is similar to the
standing waves on a fixed string. The requirement for $\chi$ to be
imaginary, bounds the quasi-resonance frequencies by the field mass
$$|\om_{QRM}|<m.$$
Eq. \p{FE} also illustrates that phenomenon of quasi-resonance exists
since the potential is not zero at spatial infinity. If the
potential is zero for $r^*\rightarrow\pm\infty$ no
quasi-resonant oscillations can exist.
Thus, for instance, in the spectrum of the  Schwarzschild-de Sitter
background \cite{Zhidenko}, no quasi-resonant oscillations can exist
even for massive scalar field.

The numerical investigation shows, that increasing of the field
mass gives rise to decreasing of the imaginary part of the QNM
until reaching the vanishing damping rate. When some threshold values
of $m$ are exceeded,  the $particular$ QN modes disappear
(see figures \ref{l=0low} and \ref{l=1low}).
We can see, that the larger field mass is, the more first overtones
share this destiny. Thus disappearing of the QNM  does not induce
the disappearing of the $whole$ QN spectrum.
Since, as it was proved in \cite{france}, the number of QN modes is
infinite for each multi-pole number, the presence of any  finite  mass
of the field  will not induce disappearing of the whole spectrum, leaving
a countable number of QN modes.

Note also, that the presence of the mass term does not change the
asymptotic crucially and, in fact, QN frequencies found under boundary
conditions \p{EB} with the 6th order WKB approximation \cite{WKB6} do not differ
from those found under the Leaver method where sub-dominant asymptotic
term is taken into account, at least for not large values of  $m$ (for
more references on the usage of WKB method for finding QNMs see for
instance \cite{WKB}).

{\bf 3.High overtones }

\begin{figure}
\centerline{\includegraphics{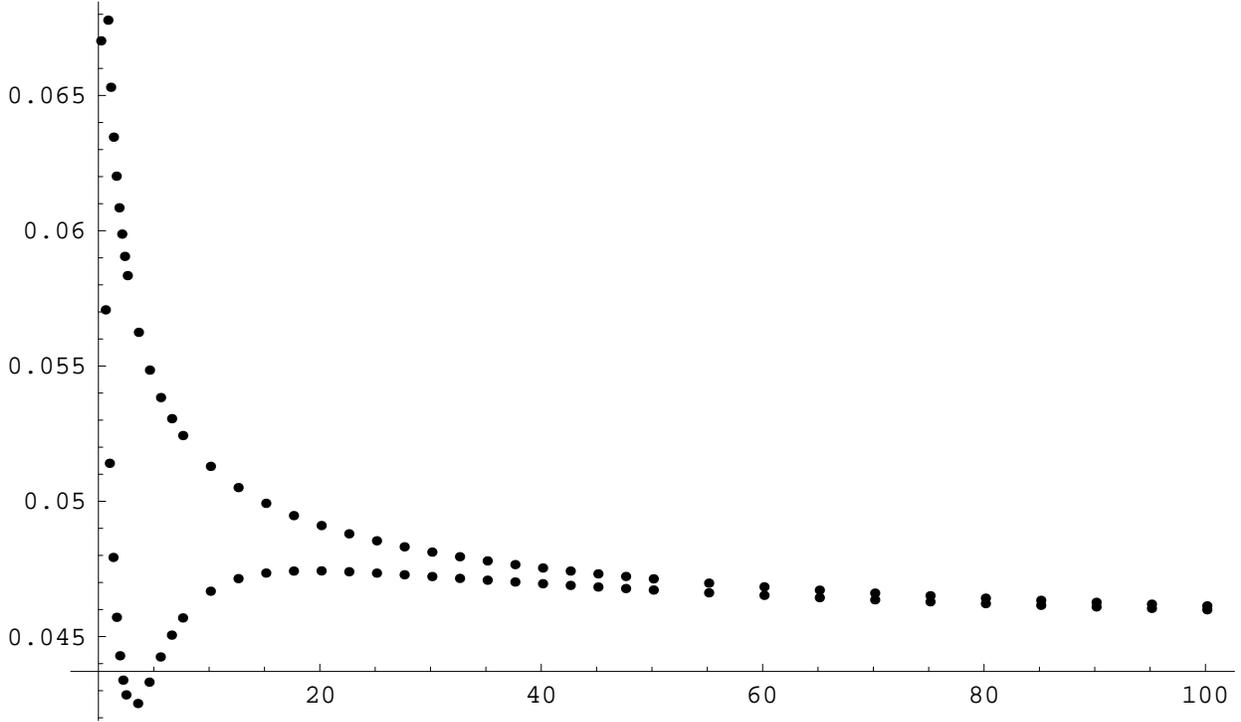}}
\caption{The real part of high overtones of $l=0$, $m=0.3$ and $m=3$
(M=1) asymptotically approaches $\frac{\ln
3}{8\pi}$}\label{l=0high}
\end{figure}
First of all, let us propose some illustrative estimations
for high-overtone behavior of the massive field.
Starting from expansion of $\chi$ at high overtones in the following
form
\be\label{chiexpand}
\chi =
\om-\frac{m^2}{2\om}+O\left(\frac{1}{\om^3}\right),
\ee
and substituting \p{chiexpand} in \p{RC}, one can find
$$
\a_n(m)=\a_n(m=0);
\quad
\b_n(m)=\b_n(m=0)-16M^2m^2+O\left(\frac{1}{\om}\right);
\quad
\g_n(m)=\g_n(m=0)+O\left(\frac{1}{\om}\right).
$$
We see that for high overtones ($M\Im{\om}\gg1$), the behavior of
coefficients is the same as in massless case with some shifted
multi-pole number $l$. Reminding that asymptotical behavior
of QNMs does not depend on the  multi-pole number we can conclude that
\be\label{HO}
\om(m,N)=\om(m=0,N)+O\left(\frac{1}{\om}\right)=\frac{N-1/2}{4M}\imo+\frac{\ln
3}{8\pi M}+O(N^{-1/2}).
\ee
If we do not pay attention to terms like $n/\om$,
this arguing illustrates why the asymptotical behavior of QNMs
does not depend on the field mass. To prove this fact $exactly$, one
should to reproduce all the steps in \cite{Motl}. We omit them
because they are almost exactly the same as in \cite{Motl}, but
require additional discussions. Finally, we find the equation for
high overtones that leads to the solution \p{HO}:
$$\tan\left(2M\imo\pi\om\right)\tan^2\left(M\imo\pi(\om+\chi)^2/\chi\right)=
\tan\left(2M\imo\pi\om\right)\tan^2\left(2M\imo\pi\om+O\left(\frac{1}{\om^3}\right)\right)
=\pm\imo+O\left(\frac{1}{\sqrt{\om}}\right).$$

Numerical calculations confirm our arguments. On figure
\ref{l=0high} you can see that the difference between modes for
$m=0.3$ and $m=3$ is small for high overtones. More accurately,
we constructed fits for both curves and found that they tend to
\p{HO}. (Fits and data for different $l$ are available from the second
author upon request).

\vspace{4mm}

{\bf Conclusion}

We have made a thorough investigation of the quasinormal spectrum for
massive scalar field in a Schwarzschild black hole
background. It is shown that the mass  of the field has
crucial influence on damping rate of the QNMs. In particular,
the greater the mass of the field is, the less the damping rate.
As a result, purely real modes appear which corresponds 
to non-damping oscillations, and,  when the field mass is greater than
some threshold values, the lower overtones disappear.
This happens however with lower overtones only,
while all the remaining higher overtones are still damping. Generally, we proved, that such
arbitrary long living modes are forbidden, unless the effective
potential is  non-zero at least at some of the two boundaries
(event horizon or spatial infinity). 
Note, that since we are working in a linear approximation and also
since the QN modes do not form a complete set, the existence of 
arbitrary long living modes, apparently, does not mean we can really have 
a system without damping. We should expect that in the
fully non-linear analysis the quasi-resonance modes may have non-vanishing 
damping rate.

In addition we proved that
at asymptotically high overtones the real part of the QNMs goes
to $\ln3/(8\pi M)$ for any (finite) mass of the field. The numerical,
analytical and WKB methods we used  show excellent agreement in
their range of validity. It is interesting for us  to find out
what will change in the above picture if considering massive fields
of higher spin. The massive Dirac field represents apparently the
most easy case.

\vspace{3mm}

{\bf Acknowledgements}

The authors acknowledge Andrei Starinets for a most useful
discussion. The work of R.K. was supported by FAPESP (Brazil).

\end{document}